\newcommand{\D}{\mathrm{d}}
\newcommand{\E}{\mathrm{e}}
\newcommand{\I}{\mathrm{i}}
\newcommand{\spread}{\renewcommand{\arraystretch}{2.0}}

\newcommand{\sech}{\mathrm{sech}}

\newcommand{\bscco}{Bi$_2$Sr$_2$CaCu$_2$O$_{8+ x}$}
\newcommand{\ybco}{YBa$_2$Cu$_3$O$_{6 + y}$}

\documentclass[aps, prb, twocolumn, showpacs,floatfix, amssymb]{revtex4}

\usepackage{graphicx}
\usepackage{dcolumn}
\usepackage{bm}
\usepackage[?]{amsmath}

\begin{document}


\title{Effective magnetic penetration depth in superconducting cylinders and spheres \\with highly anisotropic electrodynamics}
\author{D.~M.~Broun}
\author{W.~A.~Huttema}
\affiliation{Department of Physics, Simon Fraser University, Burnaby, BC, V5A 1S6, Canada}

\date{\today}

\begin{abstract}
Effective magnetic penetration depth and microwave surface impedance are derived for anisotropic layered superconductors in the shape of spheres and long cylinders, where the external magnetic field is applied in the plane of the highly conducting layers to induce out-of-plane screening currents.  The results are extended by analytic continuation to highly anisotropic conductors and to lossy superconductors at high frequency.  The electrodynamics for the general case of a superconductor or metal with arbitrary anisotropy are presented.  The treatment is then specialized to layered materials with unixaxial anisotropy, in which the penetration depth for currents flowing perpendicular to the layers, $\lambda_c$, is much greater than that for in-plane currents, $\lambda_a$.  Exact solutions are found in the limit $\lambda_a \to 0$, and are expected to provide an accurate representation of many experiments on cuprates and other layered superconductors, particularly on grain-aligned powders.
\end{abstract}

\pacs{74.25.Ha, 74.25.Nf, 74.72.-h , 74.70.Kn }
\maketitle

\section{Introduction}\label{intro}

In many superconductors of current interest the electronic structure has strongly anisotropic character and the materials can be regarded as nearly two dimensional.  Examples include the cuprate high temperature superconductors,\cite{orenstein00,bonn06} the organic superconductors,\cite{singleton02} and the recently discovered iron-based oxy-pnictides.\cite{takahashi08,chen08,yang08}  This reduced dimensionality gives rise to much interesting physics, such as enhanced fluctuation effects,\cite{lawrencedoniach} and may even be essential to the elevated transition temperatures in two dimensional materials.\cite{monthoux99} However, strong electrical anisotropy leads to difficulties in the interpretation of electrodynamic experiments, particularly at RF and microwave frequencies where, in certain important geometries, the requirement to form closed current loops creates admixtures of in-plane and out-of-plane responses.

Finite size effects, which occur when characteristic electromagnetic length scales, such as penetration depth $\lambda$ and skin depth $\delta$, become comparable to sample size, present an additional level of complexity in electrically anisostropic materials, and are often relevant in layered superconductors.  For instance, while the in-plane penetration depth $\lambda_a$  in the \ybco\ system is fairly short,  ranging from $0.1~\mu$m in optimally doped material\cite{peregbarnea04} to $2~\mu$m in heavily underdoped material,\cite{broun07} the out-of-plane penetration depth $\lambda_c$ ranges from $1~\mu$m to $100~\mu$m across the same range,\cite{peregbarnea04,hosseini04} becoming comparable to typical crystal sizes.  \bscco\ is even more electrically anisotropic, with $\lambda_c \sim 100~\mu$m at optimal doping.\cite{cooper90}  Moreover, these are low temperature values --- penetration depths grow on approach to the superconducting transition, crossing over to the much larger normal-state skin depth above $T_c$.

The study of finite-size effects in isotropic superconductors dates from the first measurements of magnetic penetration depth, and solutions have been given for a wide range of geometries.\cite{london,shoenberg}  Finite-size effects in anisotropic superconductors have been considered more recently, with various approaches  taken, depending on the geometry of the experiment.  In all cases, a component of the external magnetic field is applied \emph{parallel} to the highly conducting layers, inducing screening current loops with some component flowing \emph{perpendicular} to the layers, along the crystalline $c$~axis.  In the simplest geometries, the magnetic field is applied perpendicular to the $c$~axis, and the sample is assumed to be long in the field direction, eliminating demagnetization effects.  As we will see below, this results in a two-dimensional screening equation that is quite tractable and is solved to obtain  the internal magnetic field distribution.  Gough and Exon,\cite{gough94} and independently Mansky and co-workers,\cite{mansky94} have considered such a case for an electrically anisotropic superconductor with rectangular cross-section.  Their models provide  realistic representations of experiments on platelet crystals, which are usually thin in the $c$ direction.  Porch and Waldram, working on grain aligned powders, were the first to discuss the effective penetration depth of an electrically anisotropic sphere.\cite{waldram94,porchthesis}  In the limit that $\lambda_c$ is large compared to both $\lambda_a$ and the radius of the sphere, $a$, screening currents are weak and demagnetization effects are negligible.  Current loops, by symmetry, are then circular, and they showed that the internal field profile could be related to that of a sphere with isotropic electrodynamics and equivalent isotropic penetration depth $\lambda^\mathrm{eq} = \frac{1}{\sqrt{2}}\lambda_c$.  This particular result has found much use in the analysis of experiments on grain-aligned powders,\cite{porch93,waldram94,panagopoulos96} and is simple to apply because it does not depend on $\lambda_a$.  In the opposite limit, $\lambda_c \ll a$, Porch and Waldram argued that the current loops would again be approximately circular, and that the problem could be mapped onto an isotropic sphere with $\lambda^\mathrm{eq} = \frac{2}{\pi}\lambda_c$.  

Here we present the first complete solution to this problem, solving both cylindrical and spherical models for values of $\lambda_c$ ranging from zero to much greater than $a$.  We confirm that the limits discussed by Waldram and Porch are correct in the case of a long cylinder.  However, we now obtain exact results for the sphere and show that there are significant departures from the cylindrical case, particularly in the limit $\lambda_c < a$.  The paper is organized as follows.  In Sec.~\ref{electrodynamics} we derive the electromagnetic screening equations for an anisotropic superconductor, for both 2D and 3D geometries, and then show how dissipation at finite frequencies can be incorporated into the electrodynamics using analytic continuation.   In Sec.~\ref{cylinder} we consider the problem of a long cylinder with anisotropic electrodynamics.  This is useful both as a lead-in to the more difficult problem of the anisotropic sphere, and for demonstrating how experimentally accessible quantities such as surface impedance and penetration depth can be obtained from the internal field distribution of the superconductor.  In Sec.~\ref{sphere} we set up and solve the problem of a spherical sample in which $\lambda_c \gg \lambda_a$.  This is followed by a summary of our results in Sec.~\ref{conclusions}.

\section{Electrodynamics of anisotropic superconductors}\label{electrodynamics}

Unconventional superconductors such as the cuprates and organics are usually well described by London electrodynamics, as in most instances penetration depth is long compared to coherence length, causing the fields to vary slowly over the scale of a Cooper pair.  In the London gauge, $\bm{\nabla} \cdot \bm{A} = 0$ and the supercurrent density $\bm{j}_s$ is directly related to the vector potential $\bm{A}$ by $\bm{j}_s = - \bm{\Lambda} \cdot \bm{A}$.  Here $\bm{\Lambda}$ is the London parameter.  In general $\bm{\Lambda}$ is a tensor quantity and is what characterizes the anisotropic electrodynamic response of layered superconductors.  The screening equation that describes the Meissner state follows from combining London electrodynamics with Maxwell's equations:
\begin{eqnarray}
\bm{B} = \mu_0 \bm{H} & = & \bm{\nabla} \times \bm{A}\\
& = & - \bm{\nabla} \times \left(\bm{\Lambda}^{-1} \cdot \bm{j}_s \right)\\
& = & - \bm{\nabla} \times \left(\bm{\Lambda}^{-1} \cdot \bm{\nabla} \times \bm{H} \right)\;.\label{curlequation}
\end{eqnarray}
In the isotropic case the London parameter is a scalar, $\Lambda$, leading to the usual screening equation for magnetic field,
\begin{equation}
\nabla^2 \bm{H} = \bm{H}/\lambda^2\;,
\end{equation}
where $\lambda = (\mu_0 \Lambda)^{-1/2}$ is the London penetration depth.  In the case of anisotropic electrodynamics, and working in cartesian coordinates, the inverse London parameter is the diagonal tensor
\begin{equation}
\bm{\Lambda}^{-1} = \mu_0~\mathrm{diag}\left(\lambda_x^2,\lambda_y^2,\lambda_z^2 \right).
\end{equation}
Inserting this into Eq.~\ref{curlequation} we obtain the general form for the anisotropic screening equation,
\begin{equation}
 \left(\!\!\!\begin{array}{c}H_x \\H_y \\H_z\end{array}\!\!\!\right)
\!\! = \!\!\left(\!\!\!\spread\begin{array}{c}
 \lambda^2_z\left(\partial_{yy} H_x - \partial_{xy} H_y\right) + \lambda^2_y \left(\partial_{zz} H_x - \partial_{xz} H_z\right) \\ \lambda^2_x\left(\partial_{zz} H_y - \partial_{yz} H_z\right) + \lambda^2_z \left(\partial_{xx} H_y - \partial_{xy} H_x\right) \\ \lambda^2_y\left(\partial_{xx} H_z - \partial_{xz} H_x\right) + \lambda^2_x \left(\partial_{yy} H_z - \partial_{yz} H_y\right)\end{array}\!\!\!\right),
 \label{eqnfullscreen}
\end{equation}
where, for instance, $\partial_{xy}$ denotes the partial derivative $\frac{\partial^2}{\partial x \partial y}$, and so on.
In what follows we will focus on layered superconductors with uniaxial symmetry, in which case $\lambda_x = \lambda_y \equiv \lambda_a$ and $\lambda_z \equiv \lambda_c$.

A slight modification of the approach outlined above allows dissipative electrodynamics at finite frequencies to be treated within the same framework as the static screening response in the Meissner state.  At finite frequency, the static electromagnetic fields are replaced by harmonically varying, quasistatic fields, represented by phasor notation.  For example, 
\begin{equation}
\bm{H} \to \bm{H}(t) = \mathrm{Re}\left\{\bm{\tilde{H}} \E^{\I \omega t} \right\}\;.
\end{equation}
In this case, penetration depth $\lambda$ is replaced by a complex skin depth $\tilde{\delta}$ that is determined by the complex conductivity,  
\begin{equation}
\sigma = \sigma_1 - \I\sigma_2 = \frac{1}{\I\omega \mu_0 \tilde{\delta}^2}\;.
\end{equation}
In terms of $\tilde{\delta}$, the surface impedance is $Z_s = R_s + \I X_s = \I\omega \mu_0 \tilde{\delta}$.  In the low frequency limit, where dissipation is negligible, the London results are regained, with 
\begin{eqnarray}
\sigma & \approx & - \I\sigma_2 = \frac{1}{\I\omega \mu_0 \lambda^2}\\
Z_s & \approx & \I X_s = \I\omega \mu_0 \lambda.  
\end{eqnarray}
In the normal state, where $\sigma_s \ll \sigma_1$, $\tilde{\delta}$ can be directly related to the dc resistivity, $\rho_\mathrm{dc} = 1/\sigma_1$:
\begin{equation}
\tilde{\delta} = (1 - \I) \sqrt{\frac{\rho_\mathrm{dc}}{2 \omega \mu_0}}\;.
\end{equation}
Results will be derived in the following sections in terms of the purely static superconductive parameters $\lambda_a$ and $\lambda_c$ but, in all cases, these can be generalized by analytic continuation using the substitution $\lambda \to \tilde{\delta}$.   The results for the electrically anisotropic cylinder and sphere can therefore be applied to a much wider range of measurements and scenarios than layered superconductors in the low frequency limit, including high frequency measurements on metals.  In the case of the cylinder, the results can even be applied to magnetic materials with the substitution $\mu_0 \to \mu_r \mu_0$, where $\mu_r$ is the (complex) relative permittivity.  This extension to the magnetic case does not apply to the sphere, however, as the boundary matching procedure used in the solution implicitly assumes that the magnetic permeability of the sample is equal to that of vacuum.  Nevertheless, the method used to solve the spherical case could readily be extended to handle magnetic materials by adapting the boundary matching procedure.

\section{Electrically anisotropic superconducting cylinder}\label{cylinder}

Insight into the problem of the electrically anisotropic sphere can be obtained from solving the much simpler problem of an infinitely long anisotropic cylinder.  In this case there are no demagnetizing fields, so the surface field is equal to the applied field $H_0$.  Also, symmetry dictates that only the axial component of the magnetic field is nonzero.  The geometry of the long rectangular platelet, dealt with by Gough and Exon,\cite{gough94} also has these properties.

We define the axis of the cylinder to lie along the $x$-direction, and the crystal $c$-axis to point in the $z$ direction.  We place the origin of the coordinate system on the cylinder axis. The cylinder has radius $a$ and the location of its surface is given by $y^2 + z^2 = a^2$. With this choice of coordinates, the superconducting layers lie parallel to the $xy$-plane.  In this geometry, the screening equation, Eq.~\ref{eqnfullscreen}, reduces to
\begin{equation}
H_x(y,z) = \lambda_c^2 \frac{\partial^2}{\partial y^2} H_x + \lambda_a^2 \frac{\partial^2}{\partial z^2} H_x\;.
\end{equation}
We are interested in the limit of extreme anisotropy, $\lambda_c \gg \lambda_a$, in which to good approximation we can set $\lambda_a$ to zero.  In that limit the screening equation becomes one-dimensional:
\begin{equation}
H_x(y,z) = \lambda_c^2 \frac{\D^2}{\D y^2} H_x\;.
\end{equation}
\begin{figure} [t]
\begin{center}
\includegraphics[width= 40 mm]{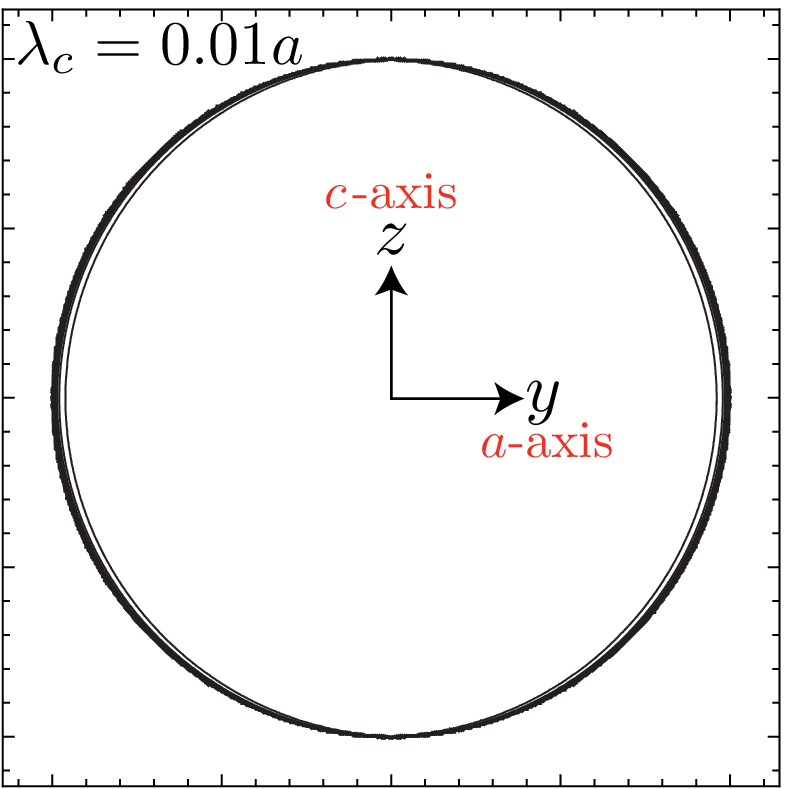}\hspace{2 mm}\includegraphics[width= 40 mm]{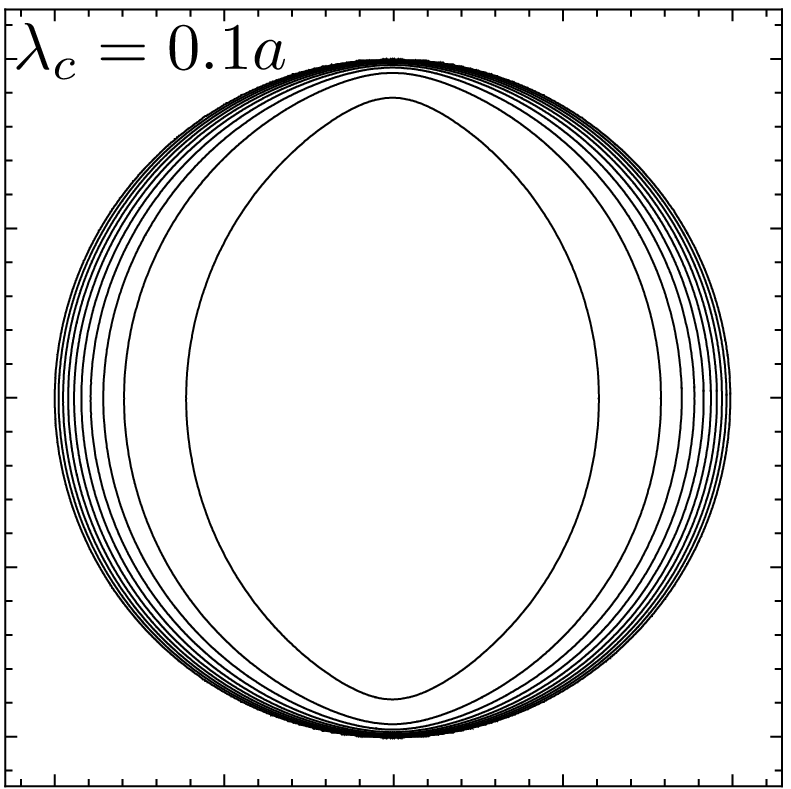}\\
\vspace{2 mm}
\includegraphics[width= 40 mm]{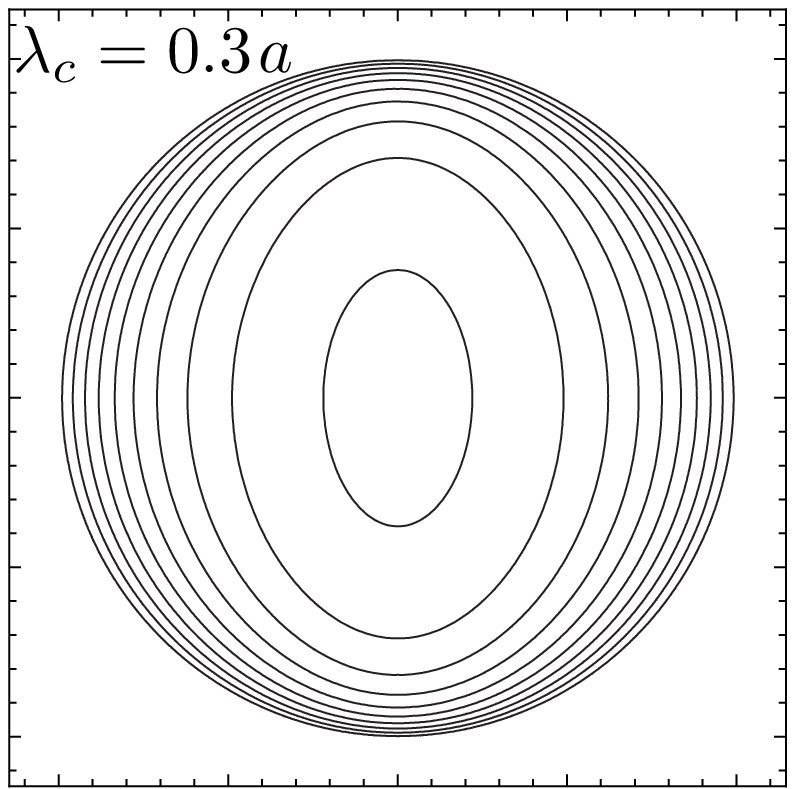}\hspace{2 mm}\includegraphics[width= 40 mm]{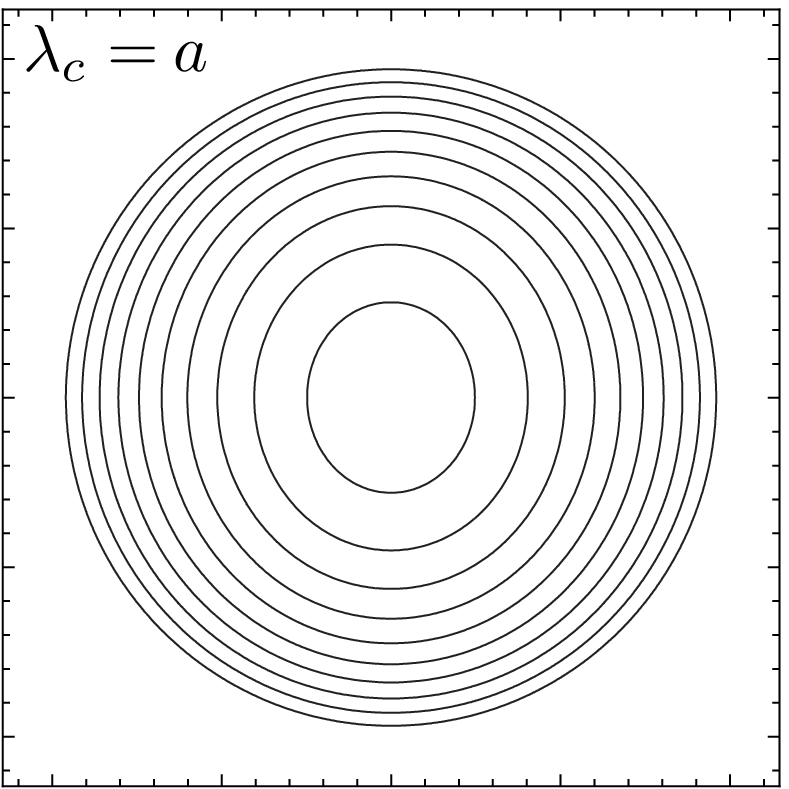}
\caption{Contours of constant field $H_x(y,z)$ inside the cylinder, for difference values of $\lambda_c/a$.  Despite the extreme anisotropy of the electrodynamics, the magnetic field distribution becomes isotropic in the limits $\lambda_c \rightarrow 0$ and $\lambda_c \gtrsim a$.}
\label{fig1}
\end{center}
\end{figure}
This limit is particularly easy to solve --- magnetic flux penetrates the sample in planes parallel to the superconducting layers, and is prevented from passing through the layers by their perfect conductivity.  We can then slice the cylinder up into a set of independent plates, parallel to the superconducting layers, and solve the screening equation separately in each one.  For each plate, the process is identical to the problem of the infinite slab of thickness $t$, the solution of which has been tabulated by Shoenberg.\cite{shoenberg}  We substitute into the slab solution, with a slab thickness that is  function of position, $t(z) = 2 \sqrt{a^2 - z^2}$, to obtain the magnetic field in the interior of the cylinder,
\begin{equation}
H_x(y,z) = H_0 \cosh \dfrac{y}{\lambda_c} \sech \dfrac{\sqrt{a^2 - z^2}}{\lambda_c}\;.
\end{equation}
Contours of contstant $H_x(y,z)$ are plotted in Fig.~\ref{fig1} for different values of $\lambda_c/a$.  Despite the extreme anisotropy of the electrodynamics, the magnetic field distribution becomes isotropic in the limits $\lambda_c \rightarrow 0$ and $\lambda_c \gtrsim a$, with the effect that screening currents will form circular loops in these two situations.

One rigorous way to calculate an effective penetration depth, $\lambda^\mathrm{eff}$, is to use the power-flow method of Gough and Exon, at finite frequency, to calculate the effective surface impedance, $Z_\mathrm{s}^\mathrm{eff} = \I \omega\mu_0\lambda^\mathrm{eff}$, corresponding to reactive power flow into the sample.\cite{gough94}  For an applied field $H_x(t) = H_0 \exp(\I \omega t)$, Faraday's law gives the average electric field:
\begin{equation}
\oint \bm{E} \cdot \D\bm{\ell} = \I \omega \mu_0 \int H_x(y,z) \D y \D z\;.
\end{equation}
The power flowing into the cylinder per unit length is
\begin{eqnarray}
P_\ell & = & \tfrac{1}{2} H_0 \E^{- \I \omega t} \oint \bm{E} \cdot \D\bm{\ell}\\
& = & \tfrac{1}{2} H_0 \E^{- \I \omega t} \I \omega \mu_0 \int H_x(y,z) \D y \D z\;.
\end{eqnarray}
The surface impedance is related to the power per unit area $P_\mathrm{A} = P_\ell/2 \pi a = \tfrac{1}{2}Z_\mathrm{s}^\mathrm{eff} H_0^2$, implying
\begin{equation}
\lambda^\mathrm{eff} = \frac{1}{2 \pi a H_0}  \int H_x(y,z) \D y \D z\;.
\end{equation}
We see that the effective penetration depth is related to the total magnetic flux inside the sample, $\Phi_B$, which can be partially evaluated in from the solution for  $H_x(y,z)$:
\begin{eqnarray}
\Phi_B & = & \mu_0 \int  H_x(y,z) \D y \D z \\ 
&= & 4 \mu_0 H_0 \int_0^a \lambda_c \tanh \dfrac{\sqrt{a^2 - z^2}}{\lambda_c} \D z\;.
\end{eqnarray}
The effective penetration depth is then
\begin{eqnarray}
\lambda^\mathrm{eff} & = & \frac{\Phi_B}{2 \pi a \mu_0 H_0}\\
& = & \frac{2 \lambda_c}{\pi a} \int_0^a \tanh \dfrac{\sqrt{a^2 - z^2}}{\lambda_c} \D z\;.
\end{eqnarray}
This is just the average of the infinite slab expression for effective penetration depth, $\lambda^\mathrm{eff} = \lambda \tanh(t/2\lambda)$,\cite{shoenberg} over one quadrant, normalized to the length of arc $\pi a/2$.  In the limit \mbox{$\lambda_c \rightarrow 0$}, $\lambda^\mathrm{eff} \rightarrow 2 \lambda_c/\pi = 0.637 \lambda_c$.  We can understand this in the following way. Screening currents in this limit are set by the total magnetic field and are confined to flow parallel to the surface.   We can therefore define an angle dependent surface impedance \mbox{$Z_\mathrm{s}(\theta) = \sqrt{\I \omega \mu_0 \rho(\theta)}$}, where $\theta$ measures the angle from the $y$ axis, with the superfluid response represented by an angle-dependent complex resistivity,
\begin{equation}
\rho(\theta) =\I \omega \mu_0 \lambda_a^2 \sin^2 \theta +  \I \omega \mu_0 \lambda_c^2 \cos^2 \theta\;.
\end{equation}
In the limit of extreme anisotropy we set $\lambda_a = 0$, then \mbox{$Z_\mathrm{s}(\theta)  =  \I \omega \mu_0 \lambda_c |\cos \theta|$}.  The effective penetration depth is the angle average of $\lambda(\theta)$: 
\begin{equation}
\lambda^\mathrm{eff}  = \langle \lambda(\theta) \rangle_\theta = \lambda_c \langle |\cos \theta |\rangle_\theta = \frac{2}{\pi}\lambda_c\;.
\end{equation}

In the opposite limit, $\lambda_c \gg a$, the effective penetration depth $\lambda^\mathrm{eff} \rightarrow a/2$.  To extract information on $\lambda_c$ in this limit it is useful to make a connection with results for the \emph{isotropic} cylinder.  The solution in the isotropic case has also been presented by Shoenberg,\cite{shoenberg} with the result that the internal field distribution is given by
\begin{equation}
H_x(r)  =  H_0 \dfrac{J_0(\I r/\lambda)}{J_0(\I a/\lambda)}\;,
\end{equation}
and the effective penetration depth is given by
\begin{equation}
\lambda^\mathrm{eff}  =  - \I \lambda \frac{J_1(\I a/\lambda)}{J_0(\I a/\lambda)}\;.
\end{equation}
Here $J_0(x)$ and $J_1(x)$ are Bessel functions of order zero and one respectively.  Since the internal field distribution of the electrically anisotropic cylinder becomes isotropic in the limit $\lambda_c \gg a$, we can make the following argument: for a given value of $\lambda_c$ in the anisotropic cylinder, there is an equivalent penetration depth $\lambda^\mathrm{eq}$ for the \emph{isotropic} cylinder that results in the same internal field distribution, and therefore the same $\lambda^\mathrm{eff}$.  The necessary mapping can be identified from series expansions of the internal fields.  For the anisotropic cylinder,
\begin{eqnarray}
H_x(y,z) & = & H_0 \cosh \dfrac{y}{\lambda_c} \sech \dfrac{\sqrt{a^2 - z^2}}{\lambda_c}\\
& = &  H_0\left(1 - \frac{a^2}{2 \lambda_c^2} + \frac{y^2}{2 \lambda_c^2} + \frac{z^2}{2 \lambda_c^2}\right) + ...\\
& = & H_0\left(1 - \frac{a^2}{2 \lambda_c^2} + \frac{r^2}{2 \lambda_c^2}\right) + ...\;,
\end{eqnarray}
where $r^2 = y^2 + z^2$.  For the isotropic cylinder,
\begin{eqnarray}
H_x(r) & = & H_0 \dfrac{J_0(\I r/\lambda)}{J_0(\I a/\lambda)}\\
 & = & H_0\left(1 + \frac{r^2}{4 \lambda^2} + ... \right)\left(1 - \frac{a^2}{4 \lambda^2} + ... \right)\\
& = & H_0\left(1 - \frac{a^2}{4 \lambda^2}+ \frac{r^2}{4 \lambda^2}\right) + ... \;,
\end{eqnarray}
where we have used $J_0(x) = 1 - x^2/4 + ... $ One expansion can be mapped on to the other with the choice \mbox{$\lambda^\mathrm{eq} = \lambda_c/\sqrt{2} = 0.707 \lambda_c$}.  To see this another way, we note that in this limit it is not the surface impedance that is angle-averaged, but the electric field.  Since the current loops are circular, the average electric field $\langle E(r)\rangle = \langle \rho(\theta)\rangle  j(r)$ is proportional to  
\begin{equation}
\lambda_a^2 \langle \sin^2 \theta\rangle+ \lambda_c^2 \langle \cos^2 \theta\rangle  = \frac{\lambda_a^2+ \lambda_c^2}{2} \rightarrow  \frac{\lambda_c^2}{2}\;.
\end{equation}

The idea of an equivalent penetration depth can be used across the full parameter range, even when the 
current loops are no longer circular.  We have shown above that as $\lambda_c \rightarrow 0$, $\lambda^\mathrm{eq} \rightarrow 2 \lambda_c/\pi$, which is not much different from the finite size limit $\lambda^\mathrm{eq} \rightarrow \lambda_c/\sqrt{2}$.  In between these limits, we can obtain the equivalent penetration depth by equating the isotropic and anisotropic expressions for the effective penetration depth $\lambda^\mathrm{eff}$:
\begin{equation}
 \lambda^\mathrm{eff} = - \I \lambda^\mathrm{eq} \frac{J_1(\I a/\lambda^\mathrm{eq})}{J_0(\I a/\lambda^\mathrm{eq})} = \frac{2 \lambda_c}{\pi a} \int_0^a \tanh \dfrac{\sqrt{a^2 - z^2}}{\lambda_c} \D z\;.
\end{equation}
This has been solved numerically, and the scale factor $\lambda^\mathrm{eq}/\lambda_c$ plotted in Fig.~\ref{LambdaEquiv} as a function of $\lambda_c/a$.

\section{Electrically anisotropic superconducting sphere}\label{sphere}

\begin{figure}[t]
\begin{center}
\includegraphics[width=60 mm]{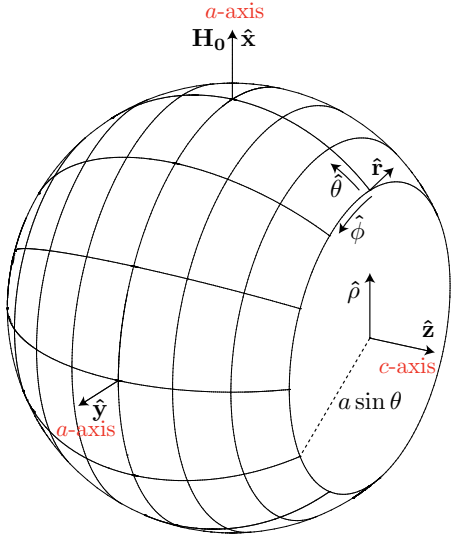}
\caption{Coordinate system for a sphere with anisotropic electrodynamics.  Magnetic field $\bm{H_0}$ is applied along the $x$ axis, in the plane of the superconducting layers. Spherical coordinates $(r,\theta,\phi)$ are used outside the sphere, and cylindrical coordinates $(\rho,\phi,z)$ inside.  This choice of geometry greatly simplifies the matching of field components at the boundary, because the azimuthal angle $\phi$ is the same inside and out.}
\label{anisosphere}
\end{center}
\end{figure}

The general case of the electrically anisotropic sphere is described by the full screening equation, Eq.~\ref{eqnfullscreen}.  To solve it, we will carry out multi-pole expansions of the interior and exterior magnetic fields and match them at the surface of the sphere.  The key to making this process tractable is to use the somewhat unusual choice of coordinate system shown in Fig.~\ref{anisosphere}, in which external field is applied along the $x$ direction and the crystal $c$ axis points is the $z$ direction, with cylindrical coordinates used inside the sphere and spherical coordinates outside. The motivation behind this choice is that the azimuthal angle $\phi$ will then be the same inside and out.

  When we specialize to the limit $\lambda_a \equiv \lambda_x = \lambda_y \rightarrow 0$, the screening equation, Eq.~\ref{eqnfullscreen}, becomes
\begin{equation}
 \left(\begin{array}{c}H_x \\H_y \\H_z\end{array}\right)
 =  \lambda^2_c \left(\begin{array}{c}
\partial_{yy} H_x - \partial_{xy} H_y \\ \partial_{xx} H_y - \partial_{xy} H_x\\ 0
\end{array}\right)\;.
\end{equation}
The infinitely strong screening response implied by \mbox{$\lambda_a \to 0$} means that magnetic flux cannot cross the superconducting layers and $H_z = 0$ inside the sphere. Combined with $\bm{\nabla}\cdot\bm{H} = 0$, we have $\partial_y H_y = -\partial_x H_x$.  The screening equation is then
\begin{equation}
 \left(\begin{array}{c}H_x \\H_y \\H_z\end{array}\right)
 =  \lambda^2_c \left(\begin{array}{c}
\partial_{xx} H_x +  \partial_{yy} H_x \\ \partial_{xx} H_y + \partial_{yy} H_y\\ 0
\end{array}\right)\;,
\end{equation}
which we will rewrite in cylindrical coordinates.  Inside the sphere the screening equation for the magnetic field decouples into separate circular slices.  We use a cylindrical coordinate system $(\rho,\phi,z)$ in which the cylinder axis lies along the $z$ direction. The azimuthal angle $\phi$ is measured from the $x$-axis.  The screening equation is $\nabla^2 \bm{H} = \bm{H}/\lambda_c^2$, with $H_z = 0$.  
Expanding this out, we have
\begin{equation}
\begin{split}
\nabla^2 \bm{H} & = \bm{\hat \rho}\left(\nabla^2 H_\rho - \frac{2}{\rho^2}\frac{\partial H_\phi}{\partial \phi} - \frac{H_\rho}{\rho^2} \right) \\
& + \bm{\hat \phi}\left(\nabla^2 H_\phi + \frac{2}{\rho^2}\frac{\partial H_\rho}{\partial \phi} - \frac{H_\phi}{\rho^2} \right)\;,
\end{split}
\end{equation}
where
\begin{equation}
\nabla^2H_\rho = \frac{1}{\rho}\frac{\partial}{\partial \rho}\left(\rho \frac{\partial H_\rho}{\partial \rho}\right) + \frac{1}{\rho^2} \frac{\partial^2 H_\rho}{\partial \phi^2}
\end{equation}
and similarly for $\nabla^2 H_\phi$. Again using $\bm{\nabla} \cdot \bm{H} = 0$ and $H_z = 0$, the azimuthal component of $\bm{H}$ can be obtained from the radial component:
\begin{equation}
\frac{\partial H_\phi}{\partial \phi} = - \frac{\partial}{\partial \rho}(\rho H_\rho)\;.
\end{equation}
The screening equation for the radial component then reduces to
\begin{equation}
 \frac{\partial^2 H_\rho}{\partial \rho^2} + \frac{3}{\rho} \frac{\partial H_\rho}{\partial \rho} + \frac{1}{\rho^2}  \frac{\partial^2 H_\rho}{\partial \phi^2}\frac{H_{\rho}}{\rho^2}+\frac{H_{\rho}}{\rho^2}= \frac{H_\rho}{\lambda_c^2}\;.
\end{equation}
Separating variables we have $H_\rho = R(\rho) \Phi(\phi)$ and
\begin{equation}
\rho^2 \frac{\;\,\,R''}{R} + 3\rho \frac{\;R'}{R} - \frac{\rho^2}{\lambda_c^2} + 1 = - \frac{\;\,\Phi''}{\Phi} = \nu^2\;.
\end{equation}
The azimuthal component has solutions of the form $\Phi(\phi) = A \cos(\nu \phi) + B \sin (\nu \phi)$, where $A$ and $B$ are constants and the $\nu$ are positive integers. The radial equation is
\begin{equation}
R''(\rho) + \frac{3}{\rho}R'(\rho) - \left(\frac{1}{\lambda_c^2} + \frac{\nu^2-1}{\rho^2} \right)R(\rho) = 0\;.
\end{equation}
Keeping only solutions that are regular at and even about the origin, the general form for the radial solution is\cite{perry}
\begin{equation}
R(\rho) = \sum_{\nu = 1}^\infty D_\nu \frac{I_\nu(\rho/\lambda_c)}{\rho}\;,
\end{equation}
where the $D_\nu$ are constants and the $I_\nu(x)$ are modified (or \emph{hyperbolic}) Bessel functions of the first kind, of order $\nu$.  Including the azimuthal dependence, the most general form for the radial component of the interior field is
\begin{equation}
H_\rho^\mathrm{int}(\rho,\phi) = \sum_{\nu = 1}^\infty \frac{I_\nu(\rho/\lambda_c)}{\rho} \big(A_\nu \cos (\nu \phi) + B_\nu \sin (\nu \phi)\big)\;.
\end{equation}
The $\sin(\nu \phi)$ terms are ruled out by the symmetry $H_\rho(-y) = H_\rho(y)$.  Also, as we will discuss below, only the $\nu = 1$ component is required to match to the exterior fields.  In this case the interior field simplifies to
\begin{equation}
H_\rho^\mathrm{int}(\rho,\phi,z = a \cos \theta) = \frac{h(\theta) a \sin\theta\;I_1(\rho/\lambda_c) \cos \phi}{\rho\;I_1(a \sin \theta/\lambda_c)}\;,
\end{equation}
where the expression has been normalized so that at the surface of each slice it takes the value \mbox{$H_\rho^\mathrm{int}(\rho = a \sin \theta,\phi,z = a \cos \theta) = h(\theta)$}.  By labeling the $z$ coordinate of each slice by the polar angle of its edge, we are anticipating that the exterior field will be expanded in spherical coordinates.   As shown above, the azimuthal component of the interior field can be obtained from radial component by differentiation to give
\begin{equation}
\begin{split}
H_\phi^\mathrm{int}&(\rho,\phi,z = a \cos \theta) \\
&= -  \frac{h(\theta) a \sin\theta\;\big(I_0(\rho/\lambda_c) + I_2(\rho/\lambda_c)\big)\sin \phi}{2 \lambda_c\;I_1(a \sin \theta/\lambda_c)}\;.
\end{split}
\end{equation}

\begin{figure}[t]
\begin{center}
\includegraphics[width= 42 mm]{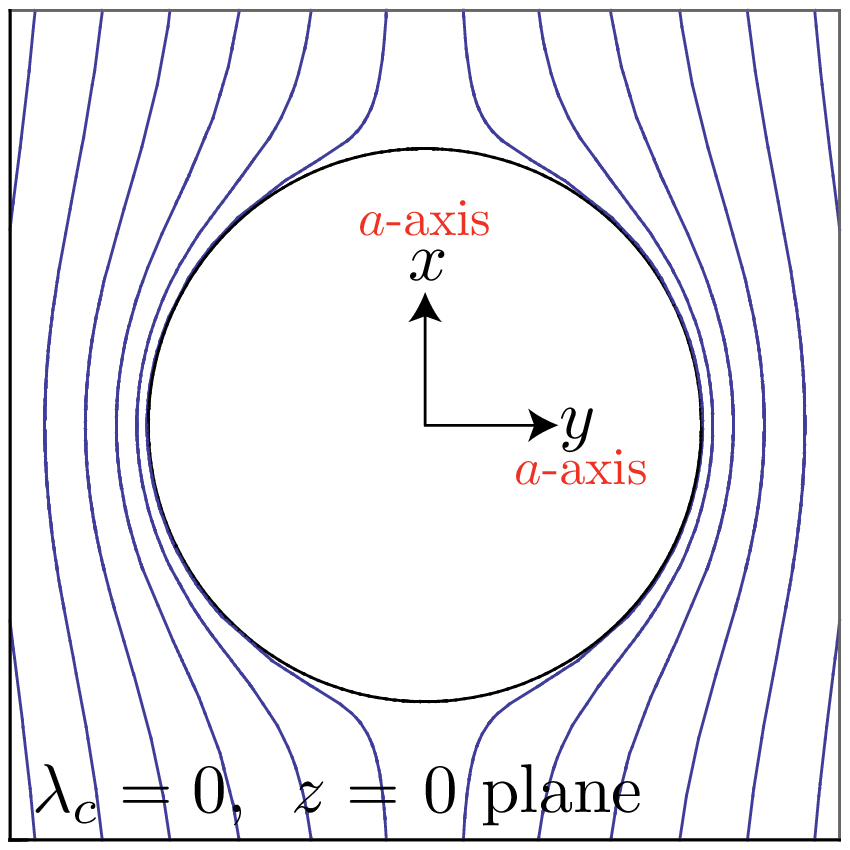}\hspace{2 mm}\includegraphics[width= 42 mm]{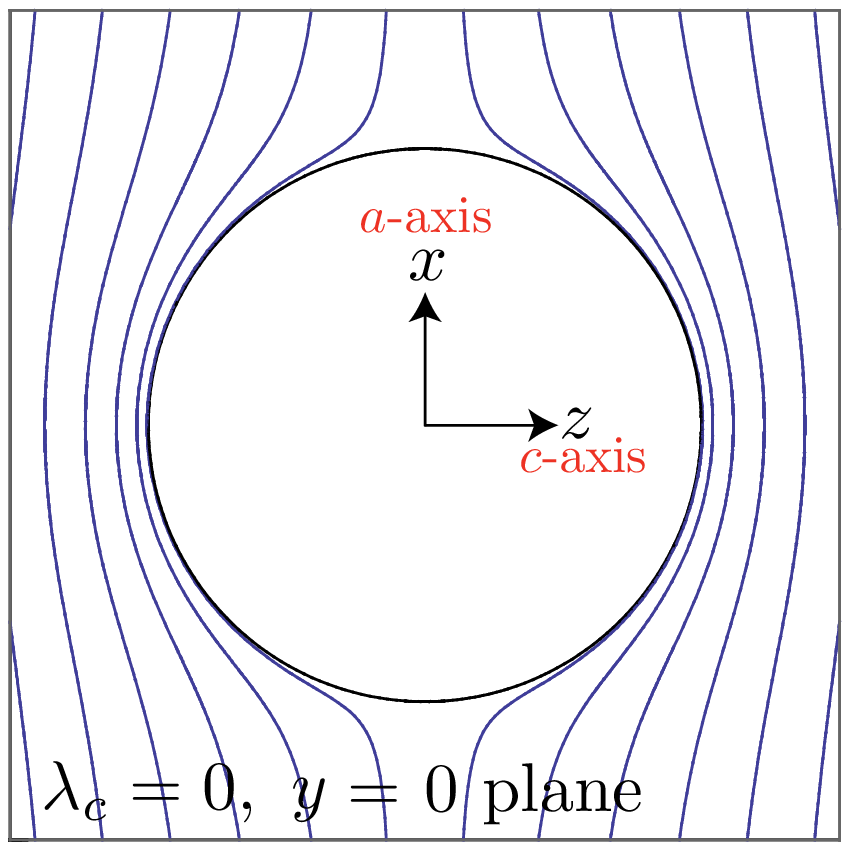}\\
\includegraphics[width= 42 mm]{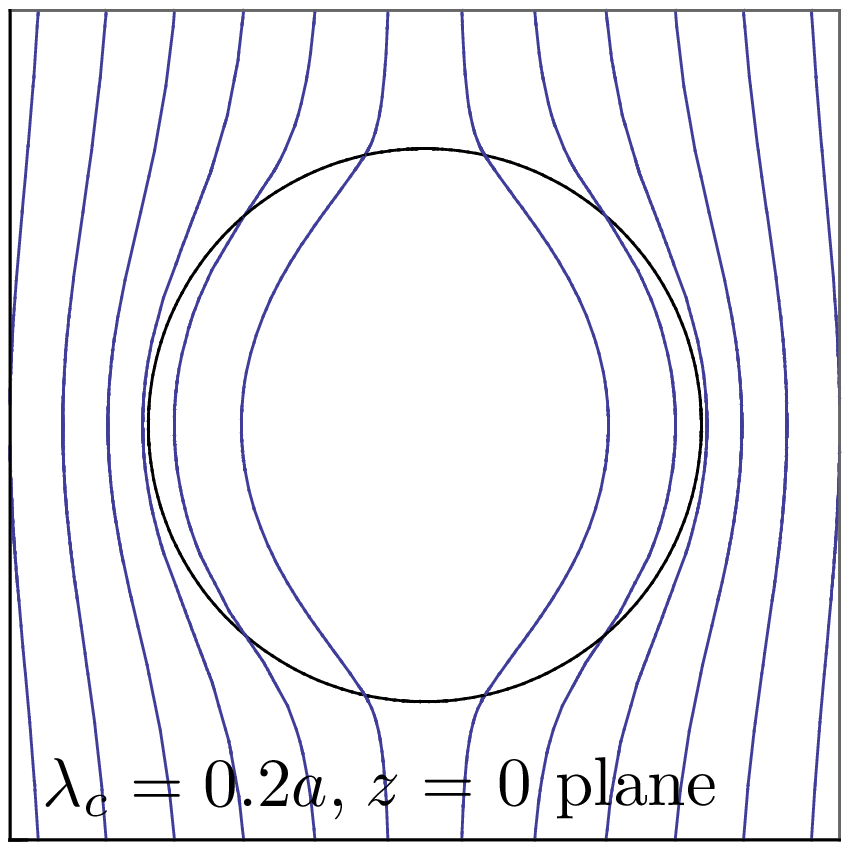}\hspace{2 mm}\includegraphics[width= 42 mm]{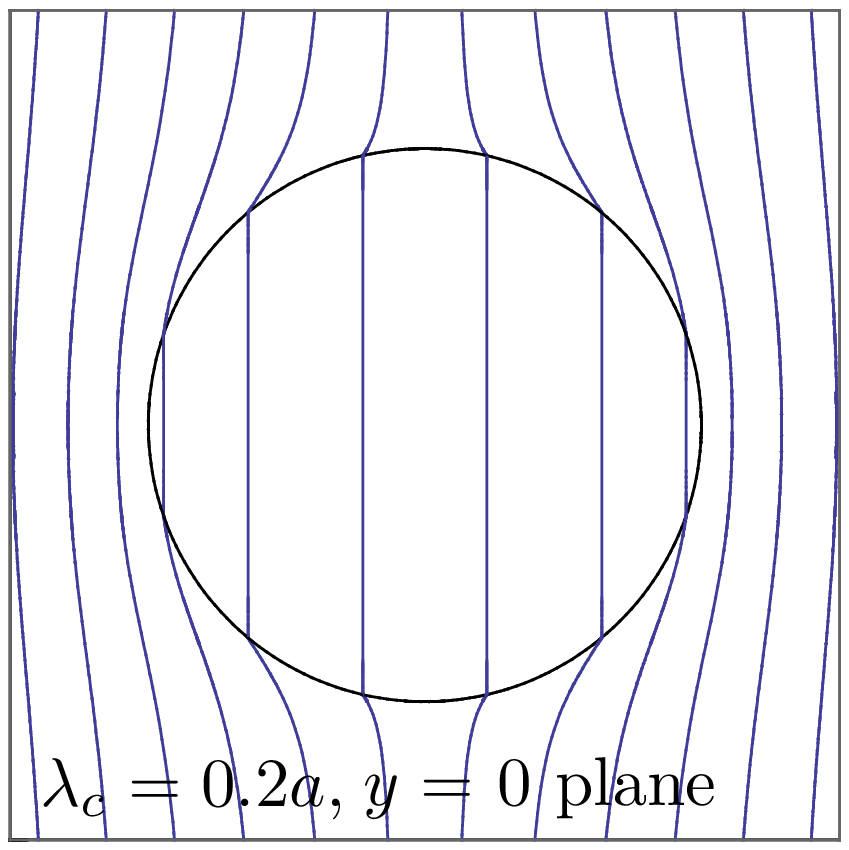}\\
\includegraphics[width= 42 mm]{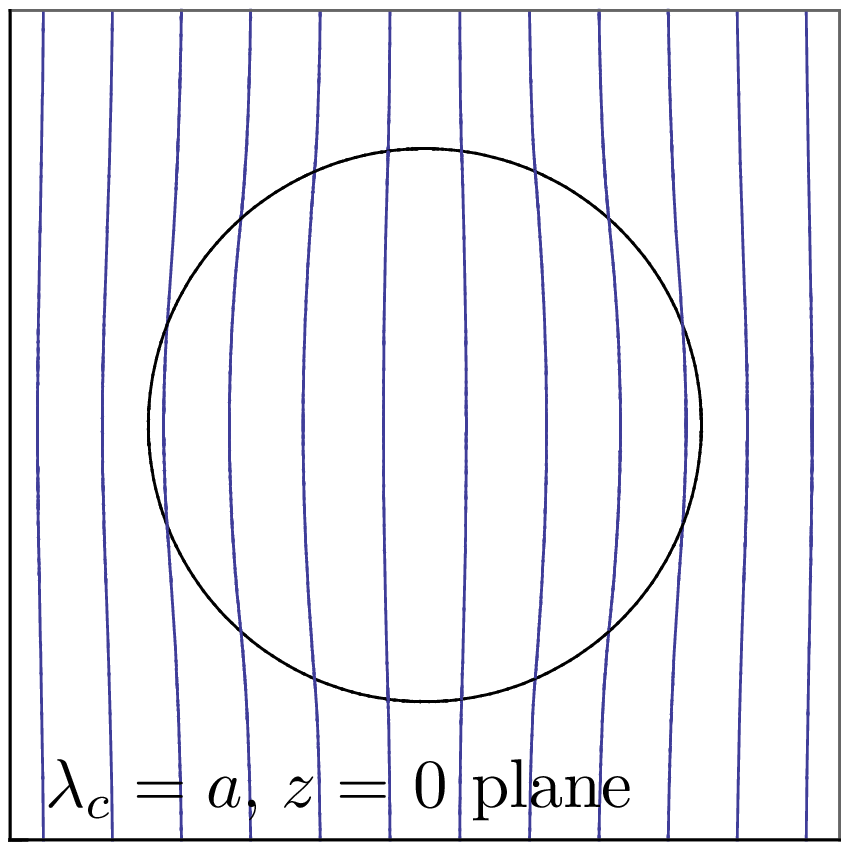}\hspace{2 mm}\includegraphics[width= 42 mm]{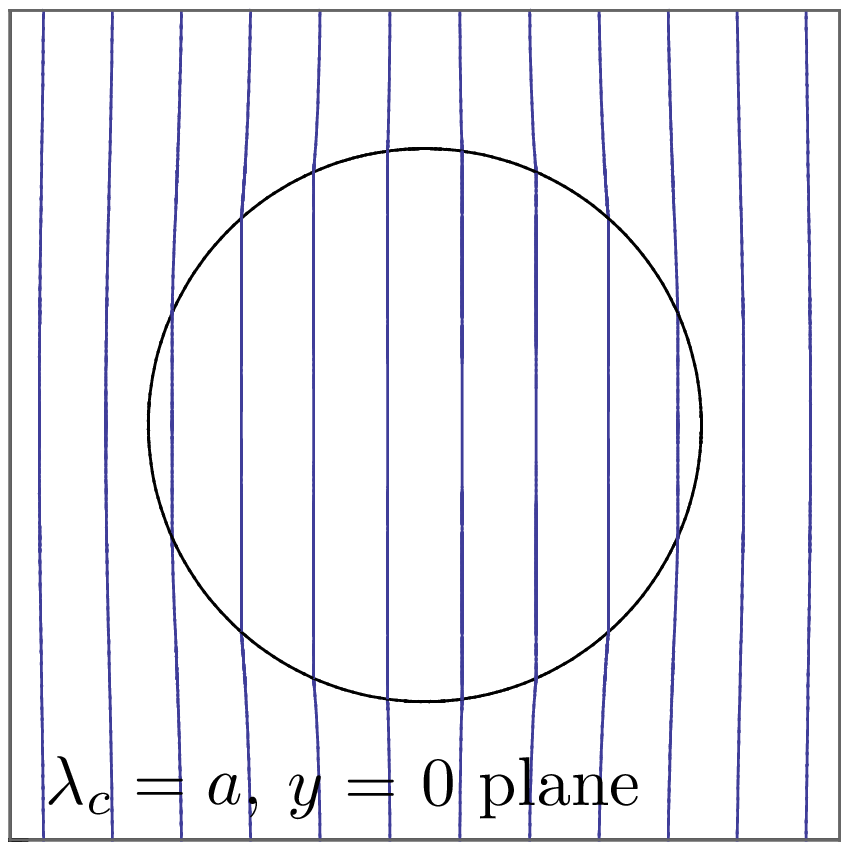}
\caption{Lines of magnetic flux show penetration of magnetic field into the electrically anisotropic sphere, as a function of $\lambda_c/a$.  Figures on the left show slices through the $z = 0$ plane, parallel to the highly conducting layers.  Figures on the right show slices through the $y = 0$ plane, in which the $c$~axis lies.  Note that flux lines cannot cross the layers, giving rise to kinks at the surface in this orientation.   $c$-axis penetration depth increases from top to bottom, with \mbox{$\lambda_c/a = 0, 0.2$} and 1 respectively.}
\label{FieldPenetration}
\end{center}
\end{figure}

For the exterior field we use spherical coordinates $(r,\theta,\phi)$, where the polar axis lies along the $z$ direction as shown in Fig.~\ref{anisosphere}. The exterior field is a solution of $\nabla^2 \bm{H} = 0$.  In the space outside the sphere there is no current, so we can obtain the magnetic field from a scalar potential, $\bm{H}^\mathrm{ext} = - \bm{\nabla} \Phi_\mathrm{m}$.
For a field $\bm{H}_0$ applied along the $x$ direction, the most general form for $\Phi_\mathrm{m}$ is
\begin{equation}
\Phi_\mathrm{m} = - H_0 r \sin \theta \cos \phi + \sum_{\ell = 1}^\infty \sum_{m = -\ell}^\ell \frac{c_\ell P_\ell^m(\cos \theta) \E^{\I m \phi}}{r^{\ell + 1}},
\end{equation}
where the $ P_\ell^m(x)$ are the associated Legendre polynomials.  The symmetry of the problem greatly restricts the number of terms in the expansion.  The symmetries of the radial component of magnetic field, $H_r(x,y,z) = H_r(r,\theta,\phi)$, are
\begin{eqnarray}
&&\!\!\!\!\!\!\!\!\!\!\!\!\begin{split}
H_r(-y) = + H_r(y) & \Rightarrow  H_r(\phi) = H_r(- \phi) \\
& \Rightarrow \cos(m \phi) \mbox{ but not } \sin(m \phi)
\end{split}\\
&&\!\!\!\!\!\!\!\!\!\!\!\!\begin{split}
H_r(-x) = - H_r(x) & \Rightarrow  H_r(\phi) = -H_r(\pi - \phi) \\
& \Rightarrow m = \mbox{odd}
\end{split}\\
&&\!\!\!\!\!\!\!\!\!\!\!\!\begin{split}
H_r(-z) = + H_r(z) & \Rightarrow H_r(\theta) = H_r(\pi - \theta) \\
&\Rightarrow \ell + m = \mbox{even} \Rightarrow \ell = \mbox{odd}\;.
\end{split}
\end{eqnarray}
In addition, only the $m=1$ term is contained in the applied field, and there is nothing in the problem to couple channels of different $m$.  For this reason we can restrict the expansion to $m=1$.  The magnetic scalar potential is then
\begin{equation}
\Phi_\mathrm{m} = - H_0 r \sin \theta \cos \phi + \sum_{\mathrm{odd}\;\ell} \frac{c_\ell P_\ell^1(\cos \theta) \cos \phi}{r^{\ell + 1}}\;.
\end{equation}
From the gradient of $\Phi_\mathrm{m}$ we obtain the three components of the exterior magnetic field:
\begin{eqnarray}
&&\!\!\!\!\!\!\!\!\!\!\!\!\begin{split}
H&_r^\mathrm{ext}(r,\theta,\phi)  \equiv  - \frac{\partial \Phi_\mathrm{m}}{\partial r} \\
&= H_0 \sin\theta \cos \phi + \sum_{\mathrm{odd}\;\ell}  \frac{c_\ell (\ell + 1) P_\ell^1(\cos \theta) \cos \phi}{r^{\ell + 2}}
\end{split}\\
&&\!\!\!\!\!\!\!\!\!\!\!\!\begin{split}
H&_\theta^\mathrm{ext}(r,\theta,\phi)  \equiv  - \frac{1}{r}\frac{\partial \Phi_\mathrm{m}}{\partial \theta} \\
&= H_0 \cos\theta \cos \phi - \sum_{\mathrm{odd}\;\ell}  \frac{c_\ell \dfrac{\D}{\D \theta}\big(P_\ell^1(\cos \theta)\big) \cos \phi}{r^{\ell + 2}}\end{split}\\
&&\!\!\!\!\!\!\!\!\!\!\!\!\begin{split}
H&_\phi^\mathrm{ext}(r,\theta,\phi)  \equiv  - \frac{1}{r \sin \theta}\frac{\partial \Phi_\mathrm{m}}{\partial \phi} \\
&= - H_0 \sin \phi + \sum_{\mathrm{odd}\;\ell}  \frac{c_\ell P_\ell^1(\cos \theta) \sin \phi}{\sin \theta\;r^{\ell + 2}}\;.
\end{split}
\end{eqnarray}
The next step is to match the fields at the surface of the sphere, $r = a$.  The azimuthal components of the interior and exterior fields both have a $\sin\phi$ angle dependence and are straightforward to match:
\begin{eqnarray}
&&H_\phi^\mathrm{ext}(r = a,\theta,\phi)  =  H_\phi^\mathrm{int}(\rho = a \sin \theta,\theta,\phi) \Rightarrow  \\
&&\begin{split}
H&_0 - \sum_{\mathrm{odd}\;\ell}  \frac{c_\ell P_\ell^1(\cos \theta)}{\sin \theta\;a^{\ell + 2}} \\
& =  \frac{h(\theta) a \sin\theta \big(I_0(a \sin \theta/\lambda_c) + I_2(a \sin \theta/\lambda_c)\big)}{2 \lambda_c I_1(a \sin \theta/\lambda_c)}\;.
\end{split}
\end{eqnarray}
Matching of the radial and polar components of the magnetic field is more complicated.  Inside the sphere, the infinite conductivity we have assumed for the CuO$_2$ layers confines the magnetic field to lie in planes parallel to the $xy$-plane, so that $H_z^\mathrm{int} = 0$.  However, if  we also assume that \mbox{$H_z^\mathrm{ext} \equiv H_r \cos \theta - H_\theta \sin \theta = 0$} just outside the sphere, we run into a contradiction.   Substituting the expansions of $H_r$ and $H_\theta$ we obtain
\begin{equation}
\begin{split}
& H_0 \sin\theta \cos \theta + \sum_{\mathrm{odd}\;\ell}  \frac{c_\ell (\ell + 1) P_\ell^1(\cos \theta) \cos \theta}{a^{\ell + 2}} \\
= & H_0 \cos\theta \sin \theta - \sum_{\mathrm{odd}\;\ell}  \frac{c_\ell \dfrac{\D}{\D \theta}\big(P_\ell^1(\cos \theta)\big) \sin \theta}{a^{\ell + 2}}\\
 = & H_0 \cos\theta \sin \theta \\& + \sum_{\mathrm{odd}\;\ell}  \frac{c_\ell \left( (\ell + 1) \cos \theta P_\ell^1(\cos \theta) - \ell P_{\ell+ 1}^1(\cos \theta) \right)}{a^{\ell + 2}}\\
& \Rightarrow  \sum_{\mathrm{odd}\;\ell}  \frac{c_\ell  \ell P_{\ell+ 1}^1(\cos \theta)}{a^{\ell + 2}} = 0\;,
\end{split}
\end{equation}
which has only the trivial solution $c_\ell = 0$, indicating that $H_z^\mathrm{ext} = 0$ only in the fully penetrated limit.  This shows that for finite $\lambda_c$ there is an abrupt change in the direction of $\bm{H}$ on crossing the surface (see plots on the right-hand side of Fig.~\ref{FieldPenetration}), implying the presence of surface currents.  This is consistent with the assumption of an infinite conductivity within the CuO$_2$ layers: by current continuity, there will in general be in-plane components of the screening current that flow at the surface.  These surface currents must have no component along the $z$ axis (the crystal \mbox{$c$ axis}) and therefore flow only in \emph{azimuthal} trajectories.  As such, they do not act as sources for $H_\phi$, showing that the azimuthal component of the field indeed remains decoupled from $H_r$ and $H_\theta$.  The matching condition for $H_r$ and $H_\theta$ is the conservation of magnetic flux.  This is equivalent to matching field components normal to the surface, requiring $H_r^\mathrm{ext}(r = a,\theta,\phi)  =  H_\rho^\mathrm{int}(\rho = a \sin \theta,\theta,\phi) \sin\theta$.  The polar components $H_\theta^\mathrm{int}$ and $H_ \theta ^\mathrm{ext}$ do not cross the surface, so do not affect flux conservation.  Their difference gives the surface current density, $\bm{j}_\mathrm{surface} = \bm{\hat r} \times \bm{\hat \theta}(H_ \theta ^\mathrm{ext} - H_ \theta ^\mathrm{int})$.

The two matching conditions at the boundary are then:
\begin{eqnarray}
&H_0 \sin\theta + \sum_{\mathrm{odd}\;\ell}  \frac{c_\ell (\ell + 1) P_\ell^1(\cos \theta)}{a^{\ell + 2}}   =  h(\theta) \sin \theta\;,\\
&\begin{split}
&H_0 \sin \theta - \sum_{\mathrm{odd}\;\ell}  \frac{c_\ell P_\ell^1(\cos \theta)}{a^{\ell + 2}} \\
& =  \frac{h(\theta) a \sin^2\theta \big(I_0(a \sin \theta/\lambda_c) + I_2(a \sin \theta/\lambda_c)\big)}{2 \lambda_c I_1(a \sin \theta/\lambda_c)}
\end{split}\\
& \hspace{-40mm} =  h(\theta) \sin \theta\;f(x)\;,
\end{eqnarray}
where $f(x)  = x(I_0(x) + I_2(x))/2 I_1(x)$ and $x = a \sin \theta/\lambda_c$. Eliminating $h(\theta)$, and defining $g(x) = f(x) - 1$, we obtain
\begin{equation}
H_0 P_1^1(\cos \theta) = \sum_{\mathrm{odd}\;\ell} \left((\ell + 1) + \frac{\ell + 2}{g(x)} \right) \frac{c_\ell P_\ell^1(\cos \theta)}{a^{\ell + 2}}\;,
\end{equation}
where we have used $P_1^1(\cos \theta) = -\sin \theta$.  Projecting onto the $P_\ell^1(\cos \theta)$ and using
\begin{equation}
\int_0^\pi P_\ell^1(\cos \theta) P_{\ell'}^1(\cos \theta) \sin \theta\;\D \theta = \frac{2 \ell (\ell + 1)}{2 \ell + 1} \delta_{\ell,\ell'}
\end{equation}
we obtain
\begin{equation}
\tfrac{4}{3} H_0 \delta_{\ell,1} = \frac{2 \ell (\ell + 1)^2}{2 \ell + 1} \frac{c_\ell}{a^{\ell + 2}} + \sum_{\mathrm{odd}\;\ell'} \frac{c_{\ell'}}{a^{\ell' + 2}} (\ell' + 2) \left\langle \ell | g^{-1}(x)| \ell'\right\rangle
\end{equation}
where
\begin{equation}
\begin{split}
& \left\langle \ell | g^{-1}(x)| \ell'\right\rangle \\
&= \frac{2 \lambda_c}{a}\int_0^\pi  \frac{P_\ell^1(\cos \theta)\;\;I_1(\frac{a \sin \theta}{\lambda_c})\;\;P_{\ell'}^1(\cos \theta)}{I_0(\frac{a \sin \theta}{\lambda_c}) + I_2(\frac{a \sin \theta}{\lambda_c}) - \dfrac{2 \lambda_c}{a \sin \theta} I_1(\frac{a \sin \theta}{\lambda_c})}  \;\D \theta\;.\label{matrixelements}
\end{split}
\end{equation}
In practice the problem is solved numerically by truncating the sum over $\ell$ at some finite $\ell = \ell_\mathrm{max}$ and solving a linear system for the $c_\ell$.  The solutions converge rapidly with $\ell_\mathrm{max}$, and in practice only 5 or so terms are required.  In the limit $\lambda_c \rightarrow \infty$, $g(x) \rightarrow 0$ and all the $c_\ell = 0$, corresponding to full field penetration.  As demagnetization effects are vanishingly small in this limit, the situation is very similar to that of the long cylinder treated above.

Another limit is which we might expect similar behaviour to the long cylinder is $\lambda_c \ll a$, but this turns out not to be the case.  To examine this limit we write
 \begin{equation}
 \begin{split}
\int_0^\pi P_\ell^1(\cos \theta)P_1^1(\cos \theta) \D \theta & = \int_0^\pi P_\ell^1(\cos \theta)(-\sin \theta) \D \theta \\
& =   \int_1^{-1} P_\ell^1(x)\D x 
\end{split}
\end{equation}
to obtain the leading order behaviour in $\lambda_c/a$:
\begin{equation}
c_\ell = H_0 a^{\ell + 2} \left(\frac{\delta_{\ell,1}}{2} - \frac{\lambda_c}{a} \frac{3(2 \ell + 1)}{4 \ell(\ell + 1)^2} \int_1^{-1} P_\ell^1(x)\D x \right)\;.
\end{equation}
The first few terms are:
\begin{eqnarray}
c_1 & = & H_0 a^3\left(\frac{1}{2} -  \frac{9 \pi}{32} \frac{\lambda_c}{a} \right)\;,\\
c_3 & = & - H_0 a^5 \frac{21 \pi}{1024}\frac{\lambda_c}{a}\;,\\
c_5 & = & - H_0 a^7 \frac{11 \pi}{2048}\frac{\lambda_c}{a}\;.
\end{eqnarray}
As $\lambda_c \to 0$, only the dipole component survives, giving
\begin{eqnarray}
H_r^\mathrm{ext} & = & H_0 \left(1 -  \frac{a^3}{r^3}\right)\sin \theta \cos \phi\;,\\
H_\theta^\mathrm{ext} & = & H_0 \left(1 +  \frac{a^3}{2 r^3}\right)\cos \theta \cos \phi\;,\\
H_\phi^\mathrm{ext} & = & - H_0 \left(1 +  \frac{a^3}{2 r^3}\right)\sin \phi\;,
\end{eqnarray}
which, when transformed back into the usual spherical coordinate system agree with the expressions for the isotropic sphere in the limit $\lambda \to 0$, as tabulated by Shoenberg.\cite{shoenberg}  However, the higher order multipole terms appear to linear order in $\lambda_c/a$ and, as we will see below, this leads to the nonintuitive result that the equivalent penetration depth of the sphere is different from that of the cylinder.

\begin{figure} [t]
\begin{center}
\includegraphics[width=80 mm]{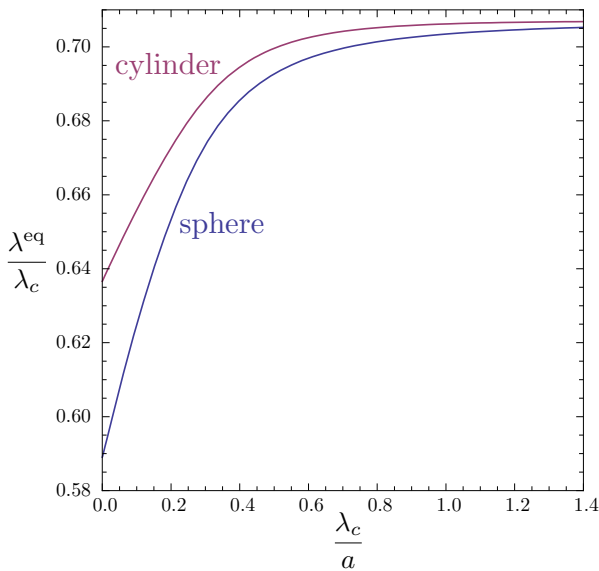}
\caption{The scale factor between the $\lambda_c$ and the equivalent isotropic penetration depth  $\lambda^\mathrm{eq}$ varies between $2/\pi=0.637$ and $1/\sqrt{2}=0.707$ for the electrically anisotropic cylinder and between $3\pi/16 = 0.589$ and $1/\sqrt{2} = 0.707$ for the electrically anisotropic sphere.} 
\label{LambdaEquiv}
\end{center}
\end{figure}

The effective penetration depth is obtained by realizing that when we carry out a penetration depth measurement, whether by microwave cavity perturbation or in a low frequency susceptibility experiment, we are effectively just measuring the dipole moment of the sample, as the higher order multipole fields fall off too rapidly with $r$ to contribute to the signal.  All information is then contained within $c_1(\lambda_c)$, allowing us to express the signal as a normalized susceptibility
\begin{eqnarray}
\frac{\chi}{\chi_0} & = & 1 - \frac{ c_1(\lambda_c)}{\frac{1}{2}H_0 a^3}\\
&  \to  & 1 - \frac{9 \pi}{16}\frac{\lambda_c}{a} \mbox{ as } \lambda_c \to 0\;.
\end{eqnarray}
This can be compared with the normalized susceptibility of the isotropic sphere, given by Shoenberg,\cite{shoenberg}
\begin{eqnarray}
\frac{\chi}{\chi_0} & = & 1 -  \frac{3 \lambda}{a}  \coth \frac{a}{\lambda} + \frac{3 (\lambda)^2}{a^2} \\
& \to & 1 - 3\frac{\lambda}{a}\mbox{ as } \lambda \to 0\;.
\end{eqnarray}
As with the cylinder, equating the results for isotropic and anisotropic cases defines $\lambda^\mathrm{eq}$, which will be a function of $\lambda_c/a$. In the limit $\lambda_c \to 0$, comparison of the leading terms gives $\lambda^\mathrm{eq} = 3 \pi/16\;\lambda_c = 0.589 \lambda_c$. The mapping is plotted at other values of $\lambda_c$ in Fig.~\ref{LambdaEquiv}, in the form of a scale factor $\lambda^\mathrm{eq}/\lambda_c$.

That the result for the sphere in the limit $\lambda_c \ll a$ is different from that of the cylinder is somewhat surprising, as the current loops in this limit are cylindrical and flow in planes parallel to the $y$--$z$ plane.  However, as pointed out above, higher order harmonics are excited to linear order in $\lambda_c/a$.  This difference was also anticipated in the approximate treatment by Waldram and co-workers,\cite{waldram94} who pointed out that the local impedance tensor on the surface of the sphere is a complicated function of angle, with principal axes that do not in general correspond with lines of latitude and longitude.

Allthough we have formulated the problem with vanishingly small in-plane penetration depth, a finite value of $\lambda_a$ could be introduced perturbatively, in the spirit of the cavity perturbation approximation used in microwave experiments.\cite{ormeno97,huttema06}  As long as $\lambda_a$ is much less than both $\lambda_c$ and the sample radius $a$, the introduction of a finite $\lambda_a$ will have a negligible effect on the internal field distribution and hence the structure of the screening currents.  Its contribution to the average surface impedance and effect penetration depth could, if required, be taken into account, for instance by carrying out a surface integral of the angle-dependent surface impedance $Z_s(\theta,\phi)$, weighted by the square of the surface magnetic field calculated in the limit $\lambda_a = 0$.  

We close this section with some comments about how to extend the results to incorporate the high frequency response of metals and lossy superconductors.  As explained in Sec.~\ref{electrodynamics}, the results for the static superconductor carry over exactly, by analytic continuation, with the replacement $\lambda_c \to \tilde{\delta}_c$. Here $\tilde{\delta}_c$ is the complex, $c$-axis skin depth and can be substituted directly into the expression for the matrix elements $\left\langle \ell | g^{-1}(x)| \ell'\right\rangle$, Eq.~\ref{matrixelements}.  The resulting linear system is then solved for the \emph{complex} dipole moment $\tilde{c}_1$, which is proportional to the measured surface impedance.  When considering the dissipative case, the scale factors plotted in Fig.~\ref{LambdaEquiv} are no longer useful.  Instead, the complex function $\tilde{c}_1(\tilde{\delta}_c)$ should be inverted numerically to obtain the complex $c$-axis conductivity directly from the measured surface impedance.

\section{Conclusions}\label{conclusions}

In conclusion, we have considered the problem of field penetration and screening in layered superconductors with highly anisotropic electrodynamics.  Exact results have been derived for samples of cylindrical and spherical shape, in 
geometries in which the external magnetic field is applied in parallel to the superconducting layers to induce screening currents with an out-of-plane component.  In contrast to approximations assumed in earlier work,\cite{waldram94,porchthesis} we find significant departures between the results for spheres and cylinders in the large-sample limit.  We have argued that when the out-of-plane penetration depth is much greater than the in-plane penetration depth, $\lambda_a$ can be set to zero without loss of accuracy.  This greatly facilitates the solution of the spherical problem, which would otherwise be intractable.  A convenient method for conveying the results is in terms of equivalent penetration depth in an isotropic system, $\lambda^\mathrm{eq}$, and data relating $\lambda^\mathrm{eq}$ to $\lambda_c$ have been tabulated for both the cylinder and the sphere.  We have also discussed how to extend the results for the static superconductor to finite frequency measurements on lossy superconductors and normal metals, using analytic continuation.

\acknowledgements

We acknowledge useful discussions with C.~P.~Bidinosti and M.~Hayden. This work was funded by the National Science and Engineering Research Council of Canada.

\end{document}